\documentclass[preprint,aps,12pt,showpacs,nofootinbib,tightenlines]{revtex4}
\usepackage{amsmath}
\usepackage{amssymb}
\usepackage{epsfig}
\usepackage{graphicx}
\begin{document}
\def\pslash{\rlap{\hspace{0.02cm}/}{p}}
\def\eslash{\rlap{\hspace{0.02cm}/}{e}}
\title{The productions of the top-pions and top-Higgs associated with the charm quark at the hadron colliders}
\author{Wenna Xu$^a$}
\author{Xuelei Wang$^{b,a}$} \email{wangxuelei@sina.com}
\author{Zhen-jun Xiao$^a$} \email{xiaozhenjun@njnu.edu.cn}
\affiliation{a: Department of Physics and Institute of
Theoretical Physics, Nanjing Normal University, Nanjing, Jiangsu 210097, P.R. China}
\affiliation{b: College of Physics and Information Engineering,
Henan Normal University, Xinxiang, Henan 453007. P.R. China}
\date{\today}
\begin{abstract}
In the topcolor-assistant technicolor (TC2) model,
the typical physical particles, top-pions and top-Higgs, are
predicted and the existence of these particles could be regarded
as the robust evidence of the model. These particles are
accessible at the Tevatron and LHC, and furthermore the
flavor-changing(FC) feature of the TC2 model can provide us a
unique chance to probe them. In this paper, we study some
interesting FC production processes of top-pions and top-Higgs at
the Tevatron and LHC, i.e., $c\Pi_{t}^{-}$ and
$c\Pi_{t}^{0}(h_{t}^{0})$ productions. We find that the light
charged top-pions are not favorable by the Tevatron experiments
and the Tevatron has a little capability to probe neutral top-pion
and top-Higgs via these FC production processes.
At the LHC, however, the cross section can reach the level of $10\sim
100$ pb for $c\Pi_t^-$ production and $ 10\sim 100$ fb for $c\Pi_t^0(h_t^0)$ production.
So one can expect that enough signals could be produced at the LHC experiments.
Furthermore, the SM background should be clean due to the FC feature of the processes
and the FC decay modes $\Pi_t^-\rightarrow b\bar{c},~\Pi_t^0(h_t^0)\rightarrow t\bar{c}$ can provide us the
typical signal to detect the top-pions and top-Higgs. Therefore,
it is hopeful to find the signal of top-pions and top-Higgs with
the running of the LHC via these FC processes.
\end{abstract}


\maketitle


\section{ Introduction}

The upgraded $p\bar{p}$ collider Tevatron is now engaged in RUN
II, and followed by the forthcoming Large-Hadonic-Collider (LHC)
with the center of mass(c.m.) energy of 14 TeV. One of the
important tasks of these hadron colliders is to detect the signal
of the physics beyond the standard model(SM). With the running of
the LHC, it is possible to find the signal of the new heavy
Higgs-like particles in the new physics models related to the
electroweak symmetry breaking(EWSB). Therefore, the LHC will open
a wide window to test the new physics models, furthermore, explore
the EWSB.

Among the various new physics theories, techinicolor(TC) model
introduced by Weinberg and Susskind\cite{TC}, offers a new
possible mechanism of the EWSB and solves the problems of the SM
neatly. As one of the promising candidates of the new physics, TC
theory has been developed for many years. In 1990s, a new
dynamical topcolor was introduced to combine with the TC theory,
then we arrived the topcolor assisted technicolor(TC2)
model\cite{TC2}. The TC2 model predicts three CP odd Pseudo
Goldstone bosons(PGBs) called
top-pions($\Pi_{t}^{\pm},\Pi_{t}^{0}$) and a CP even scalar called
top-Higgs($h_{t}^{0}$) in a few hundred GeV region. The existence
of these physical particles can be regarded as the typical feature
of the TC2 model and the observation of them is a robust evidence
of the model. Thus the study of the production processes of these
typical particles is a very interesting research work, and a lot
of studies about this aspect have been done\cite{pro-LC1, pro-LC2,
pro-LHC,proFC-LC}.¡¡Another feature of the TC2 model is that the
topcolor interaction is non-universal, and the
Glashow-Lliopoulos-Maiani(GIM) symmetry is violated which results
in the significant tree-level flavor-changing(FC) couplings. This
is an essential feature of these models due to the need to single
out the top quark for condensation. It is known that the existence
of the GIM makes the FC processes in the SM to be hardly detected,
and hence the FC processes would open an ideal window to probe the
TC2 model. Some FC processes in the TC2 model have been studied
\cite{proFC-LC, He,TCV-TC2, Burdman,Cao,TZ-TC2, TQ-TC2}. On the
other hand, the tree-level FC couplings in the TC2 model can also
result in the loop-level FC couplings: $tcZ, tc\gamma, tcg$. The
contributions of these one-loop FC couplings are also significant
which makes the rare top quark decays\cite{TCV-TC2}, $t\bar{c}$
production \cite{Burdman,Cao}, and $ tZ(\gamma)$
productions\cite{TZ-TC2} become detectable at the future LHC and
ILC. Due to the existence of the FC couplings in the TC2 model,
the top-pions and top-Higgs can be produced via some FC processes.
These FC production processes would play a very important role in
probing these new physical particles because the SM background
would be very clean for these FC processes. We have studied the FC
 production processes, $t\bar{c}\Pi_t^0$, via $e^+e^-$ and
$\gamma\gamma$ collision\cite{proFC-LC} and find that the cross
sections are large enough and the backgrounds are very small. So
these FC production processes will provide us a good chance to
search for the neutral top-pion at the planed ILC. With the
running of the LHC in 2007, the signal of the TC2 model would be
first observed in the LHC. As we know, backgrounds at the hadron
colliders are much larger than those at linear colliders which
makes the detection of new particles at the hadron colliders
become more difficult. But we find that there also exist the FC
production processes of the top-pions and top-Higgs at the hadron
colliders, i.e., the $c\Pi_t^-,~ c\Pi_t^0(h_t^0)$ productions. In
this paper, we study the potential to discover the top-pions and
top-Higgs via these FC production modes.

 We organize the rest parts of the paper as follow. In section 2, we present our
calculations  and the numerical results of the cross sections. The
conclusions are given in section 3.

\section{ Calculations and numerical results }
\hspace{1mm}

 Since the topcolor interaction treats the third
family quark differently from the first and the second families,
the TC2 model does not possess the GIM mechanism, and the
non-universal gauge interaction results in the significant
tree-level FC couplings of the top-pions(top-Higgs) to the quark
pair when one writes the interactions in the quark mass
eigen-basis. The couplings of the top-pions and top-Higgs to the
quarks can be written as\cite{He,TCV-TC2}

\begin{eqnarray}
 \cal L&=&\frac{m_{t}}{\upsilon_{w}}
 \tan\beta
 [iK_{UR}^{tt}K_{UL}^{tt*}\overline{t_{L}}t_{R}\Pi_{t}^{0}
 +\sqrt{2}K_{UR}^{tt*}K_{DL}^{bb}\overline{t_{R}}b_{L}\Pi_{t}^{+}  \nonumber\\
 &&\hspace*{1.5cm}+iK_{UR}^{tc}K_{UL}^{tt*}\overline{t_{L}}c_{R}\Pi_{t}^{0}
 +\sqrt{2}K_{UR}^{tc*}K_{DL}^{bb}\overline{c_{R}}b_{L}\Pi_{t}^{+}   \nonumber\\
 &&\hspace*{1.5cm}+i\frac{m_{b}^{*}}{m_{t}}\overline{b_{L}}b_{R}\Pi_{t}^{0}
+K_{UR}^{tt}K_{UL}^{tt*}\overline{t_L}t_Rh_{t}^{0}+K_{UR}^{tc}K_{UL}^{tt*}\overline{t_{L}}c_{R}h_{t}^{0}
 +h.c.].
 \end{eqnarray}
Where $\tan\beta=\sqrt{(\upsilon_{w}/\upsilon_{t})^2 -1}$,
\hspace{0.5cm}$\upsilon_{t}\approx 60-100$ GeV is the top-pion
decay constant, $\upsilon_{w}=246$ GeV is the EWSB scale,
$K^{ij}_{U,D}$ are the matrix elements of the unitary matrix
$K_{U,D}$, from which the Cabibbo-Kobayashi-Maskawa (CKM) matrix
can be derived as $V=K^{-1}_{UL}K_{DL}$. Their values can be
written as
 \begin{eqnarray*}
  \hspace*{1cm}K^{tt}_{UL}=K^{bb}_{DL}\approx 1,
  \hspace{1.5cm}K^{tt}_{UR}=1-\varepsilon,
  \hspace*{1.5cm}K^{tc}_{UR}=\sqrt{2\varepsilon-\varepsilon^{2}}.
  \end{eqnarray*}
$\varepsilon$ is a model dependent parameter which is in the range
of $0.03\leq \varepsilon \leq 0.1$\cite{TC2}. The mass $m_{b}^{*}$
is a part of b-quark mass which is induced by the instanton and
can be estimated as \cite{TC2}
\begin{eqnarray*}
  \hspace*{1cm}m_{b}^{*}=\frac{3\kappa m_{t}}{8\pi^{2}}\sim
  6.6\kappa GeV,
  \end{eqnarray*}
where we generally expect $\kappa\sim1$ to $10^{-1}$ as in QCD.

\begin{figure}
\begin{center}
\includegraphics [scale=0.7] {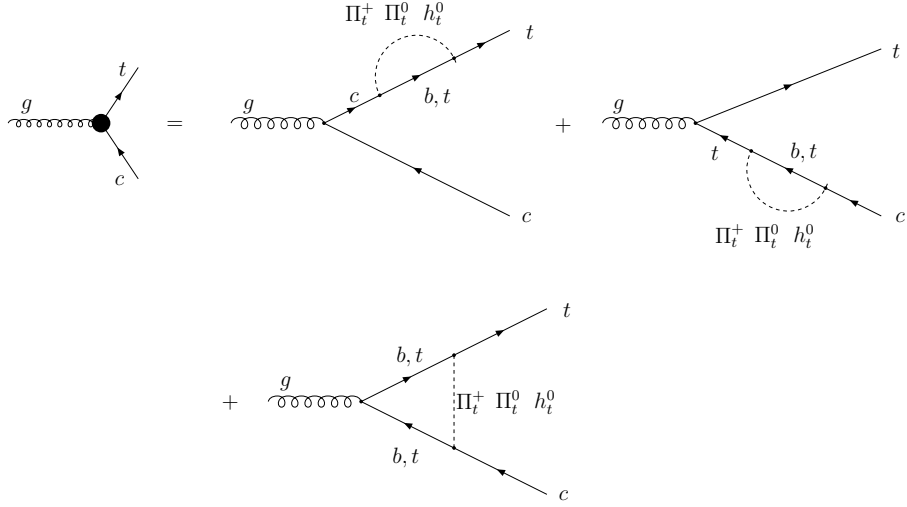}
\caption{The Feynman diagrams of the one-loop FC coupling $tcg$.}
\label{fig:fig1}
\end{center}
\end{figure}
The existence of tree-level FC couplings $\Pi_{t}^{0}t\bar{c}$,
$\Pi_t^-b\bar{c}$ and $h_{t}^{0}t\bar{c}$ can also result in the
loop-level FC coupling $tcg$ as shown in Fig.1. Because there is
no corresponding tree-level $tcg$ coupling to absorb these
divergences, the divergences just cancel each other and the total
result is finite as it should be. As we have mentioned in the
introduction, these tree-level and loop-level FC couplings can
make the significant contribution to some processes. These FC
couplings can also induce the interesting FC production processes
of top-pions and top-Higgs at the hadron colliders. i.e.,
$p\overline{p}(pp)\rightarrow c \Pi_{t}^{-}, ~c
\Pi_{t}^{0}(h_{t}^{0})$. In the following, we focus on studying
these processes.

\subsection{The $c \Pi_{t}^{-}$ production at the hadron colliders}

 We know that there is only one neutral scalar Higgs
in the SM, and hence the existence of physical charged
(pseudo)scalars can be regarded as an unambiguous signal beyond
the SM. The Tevatron can probe the charged (pseudo)scalars mass up
to $300\sim 350$ GeV, and the LHC can probe the mass-range of
charged (pseudo)scalars up to $\sim O(1)$ TeV\cite{He}. As it is
known, the charged top-pions are predicted in the TC2 model. The
FC coupling $\Pi_t^-b\bar{c}$ can result in the tree-level
production mode $c\Pi_t^-$ via $bg$ collision. On the other hand,
the loop-level FC coupling $tcg$ can also make the contribution to
the $c\Pi_t^-$ production. The Feynman diagrams of the
$c\Pi_{t}^{-}$ production at the hadron colliders are shown in
Fig.2(A-C).

\begin{figure}
\vspace{-7cm}
\includegraphics [scale=0.7] {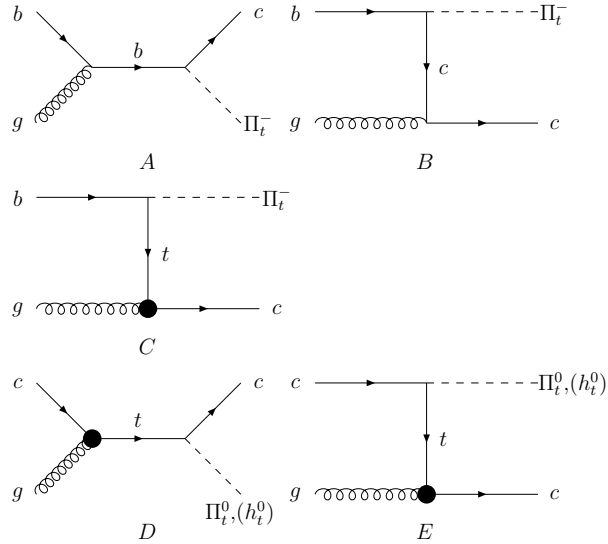}
\vspace{-65mm} \caption{ The Feynman diagrams of the $c
\Pi_{t}^{-},~c \Pi_{t}^{0}(h_t^0)$ productions at the hadron
colliders.} \label{fig:fig2}
\end{figure}

The production amplitudes are expressed as follow
\begin{eqnarray}
M_{A}&=&\frac{-i\sqrt{2}m_{t}tan\beta}{\upsilon_{w}}\sqrt{2\varepsilon-\varepsilon^{2}}g_{s}T^{a}_{ij}
G(p_{1}+p_{2},m_{b})\nonumber\\
&& \cdot \overline{u}^i_{c}(p_{3})L(\pslash_{1}+\pslash_{2}+m_{b})\eslash^a(p_{2})u^j_{b}(p_{1}),
\end{eqnarray}
\begin{eqnarray}
M_{B}&=&\frac{-i\sqrt{2}m_{t}tan\beta}{\upsilon_{w}}\sqrt{2\varepsilon-\varepsilon^{2}}g_{s}T^{a}_{ij}G(p_{3}-p_{2},m_{c})
\nonumber\\
&& \cdot \overline{u}^i_{c}(p_{3})\eslash^a(p_{2})(\pslash_{3}-\pslash_{2}+m_{c})Lu^j_{b}(p_{1}),
\end{eqnarray}

\begin{eqnarray}
\hspace{0.8cm}M_{C}&=&\frac{-i\sqrt{2}m_{t}^{3}tan^{3}\beta}{16\pi^{2}\upsilon_{w}^{3}}
(1-\varepsilon)^{2}\sqrt{2\varepsilon-\varepsilon^{2}}g_{s}T^{a}_{ij}G(p_{3}-p_{2},m_{t})\nonumber\\
&&\{[2B_{1}(p_{2}-p_{3},m_{b},M_{\Pi_{t}})+B_{1}(p_{2}-p_{3},m_{t},M_{\Pi_{t}})
+B_{1}(p_{2}-p_{3},m_{t},M_{h_{t}})\nonumber\\
&&+B_{0}(p_{2}-p_{3},m_{t},M_{\Pi_{t}})-B_{0}(p_{2}-p_{3},m_{t},M_{h_{t}})\nonumber\\
&&-B_{0}(-p_{3},m_{t},M_{\Pi_{t}})+B_{0}(-p_{3},m_{t},M_{h_{t}}) \nonumber\\
&&-m_{t}^{2}(C'_{0}+C^*_{0})-2m_{b}^{2}C_{0}+4C_{24}+2C'_{24}+2C_{24}^{*}]
\cdot\overline{u}^i_{c}(p_{3})\eslash^{a}(p_{2})(\pslash_{3}-\pslash_{2})Lu^j_{b}(p_{1})\nonumber\\
&&+[2(C_{23}+C_{12})+C'_{23}+C'_{12}+C^{*}_{23}+C^{*}_{12}]
\cdot\overline{u}^i_{c}(p_{3})\pslash_{2}\eslash^{a}(p_{2})\pslash_{3}(\pslash_{3}-\pslash_{2})Lu^j_{b}(p_{1})\nonumber\\
&&+m_{t}^{2}(C'_{0}-C^*_{0})\cdot\overline{u}^i_{c}(p_{3})\pslash_{2}\eslash^{a}(p_{2})Lu^j_{b}(p_{1})\nonumber\\
&&+m_{t}^{2}(-C'_{12}+C^*_{12})\cdot\overline{u}^i_{c}(p_{3})\eslash^{a}(p_{2})\pslash_{3}Lu^j_{b}(p_{1})\}.
\end{eqnarray}
Where, $G(p,m)=1/(p^{2}-m^{2})$ is the propagator of the particle,
$L=(1-\gamma_5)/2$. In the production amplitude $M_C$, the
three-point standard functions are defined as
$$C_{ij}=C_{ij}(p_{2},-p_{3},m_{b},m_{b},M_{\Pi_{t}}),$$
$$C'_{ij}=C_{ij}(p_{2},-p_{3},m_{t},m_{t},M_{\Pi_{t}}),$$
$$C^*_{ij}=C_{ij}(p_{2},-p_{3},m_{t},m_{t},M_{h_{t}}).$$
Here, we have ignored the mass difference between the charged
top-pions and the neutral top-pion.

\begin{figure}[tb]
 \centerline{\mbox{\epsfxsize=8cm\epsffile{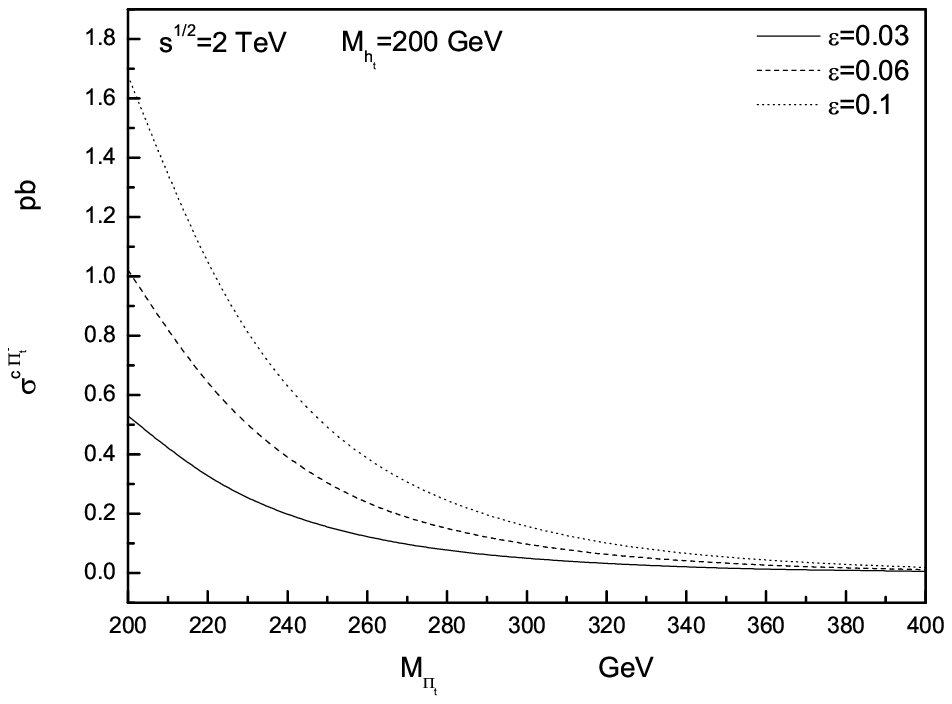}
 \epsfxsize=8cm\epsffile{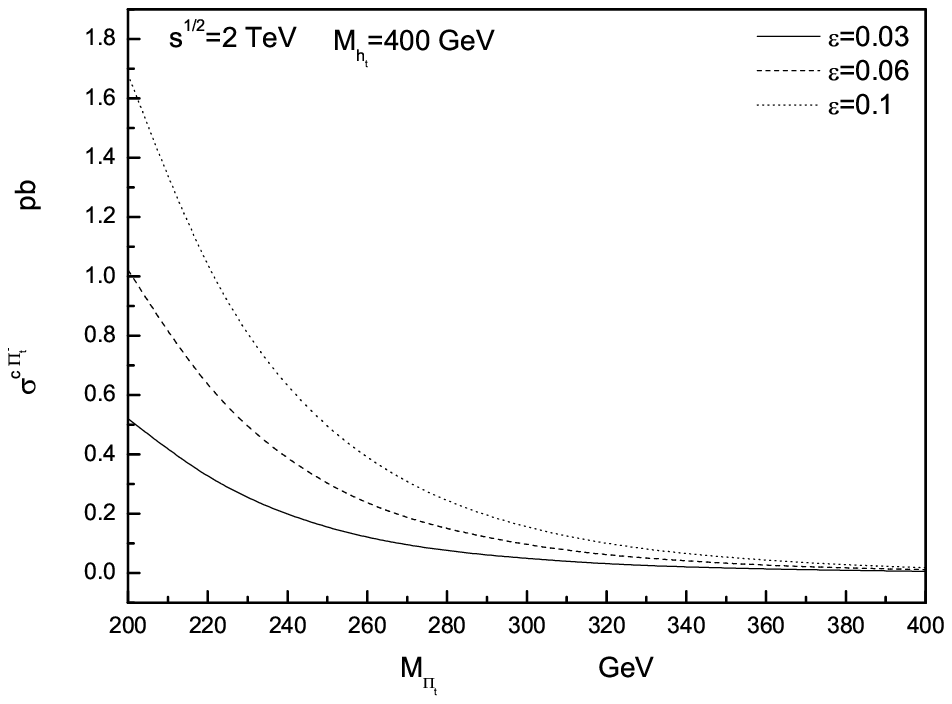}}}
\caption{The hadronic cross section of
$c\Pi_{t}^{-}$ production as a function of $M_{\Pi_t}$ at the
Tevatron, with $M_{h_t}=200, 400$ GeV, respectively.}
\label{fig:fig3}
\end{figure}

\begin{figure}[tb]
 \centerline{\mbox{\epsfxsize=8cm\epsffile{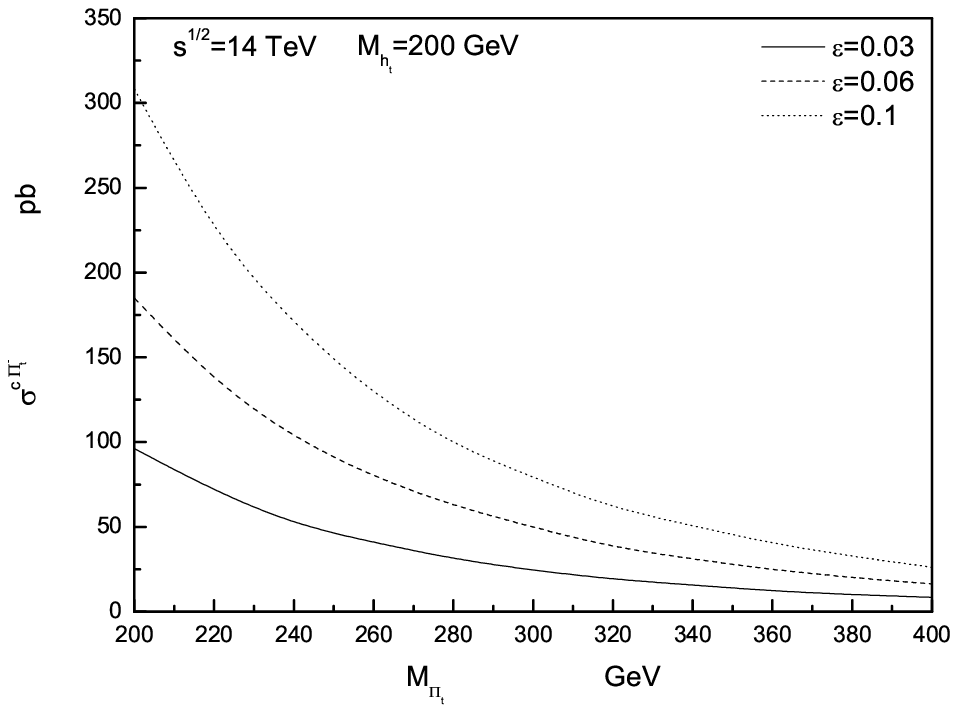}
 \epsfxsize=8cm\epsffile{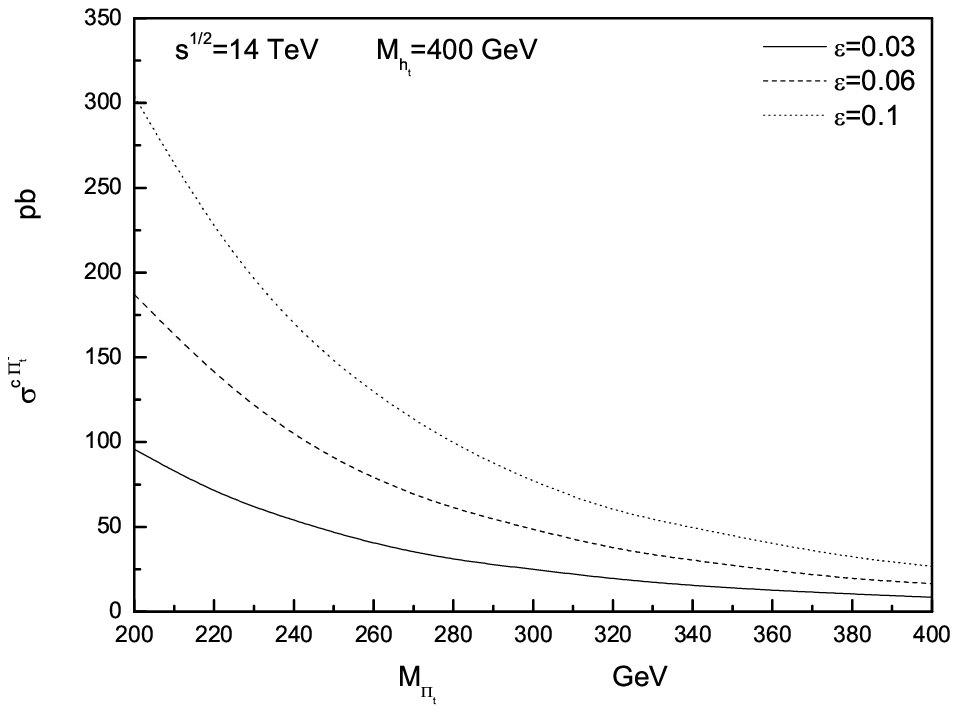}}}
\caption{The hadronic cross section of
$c\Pi_{t}^{-}$ production as a function of $M_{\Pi_t}$ at the LHC,
with $M_{h_t}=200, 400$ GeV, respectively.} \label{fig:fig4}
\end{figure}

With the above production amplitudes, we can directly obtain the
cross section of subprocess $bg\rightarrow c\Pi_t^-$. The hadronic
cross section at the Hadron colliders can be obtained by folding
the cross section of subprocesses with the parton distribution.

 To obtain the numerical results of the cross section, we take
$m_{t}=174$ GeV, $m_{b}=4.7$ GeV, $m_{c}=1.25$ GeV\cite{data},
$\upsilon_{w}=246$ GeV, $\upsilon_{t}=60$ GeV. The strong coupling
constant $g_{s}=2\sqrt{\pi\alpha_{s}}$ can be obtained from the
one-loop evolution formula at the energy of the Tevatron and the
LHC, respectively. There are three free parameters involved in the
production amplitudes: the top-pion mass $M_{\Pi_t}$(We have
ignored the mass difference between the neutral top-pion and
charged top-pions), the top-Higgs mass $M_{h_{t}}$ and the
parameter $\varepsilon$. In order to see the influence of these
parameters on the cross section, we take $M_{\Pi_t}$ to vary in
the range 200 GeV$\leq M_{\Pi_t}\leq400$ GeV, $\varepsilon=0.03,
0.06,0.1$, and $M_{h_{t}}=200,400$ GeV, respectively.

The numerical results of the cross section for the $c\Pi_t^-$
production at the Tevatron and the LHC are shown in Fig.3 and
Fig.4, respectively. The cross section should be sensitive to the
mass of final state $\Pi_t^-$, and in the Fig.3-4, we plot the
hadronic cross section as a function of $M_{\Pi_t}$. From these
figures, we can see that the production cross section decrease
sharply as $M_{\Pi_t}$ increasing because the large mass of the
top-pion can strongly depress the phase space. For the $c\Pi_t^-$
production, the top-Higgs only make a virtual contribution to the
loop-level coupling $tcg$, so the cross section is insensitive to
the mass of the top-Higgs. The dependence of the cross section on
the $\varepsilon$ is obvious, and when $\varepsilon$ becomes large
the cross section increases. As it is know, at the Tevatron, the
main contribution comes from the light quarks, so the cross
section of $c\Pi_t^-$ production is not so large in most parameter
space and the cross section can reach the level of pb only in a
narrow range for light top-pion. Because there is no clue of
existence of the charged top-pions at the Tevatron, the light
charged top-pions are not favorable by the Tevatron experiments.
At the LHC, the gluon makes the main contribution and the cross
section is greatly increased. Via the FC  production mode
$c\Pi_t^-$, a large number of signals would be produced in a wide
range of the parameter space at the LHC. So the LHC might provide
a good chance to probe the charged top-pions. On the other hand,
the main contribution comes from the tree-level figures of
Fig.2(A-B), so such charged top-pion production mode also provide
us a unique way to study the FC coupling $\Pi_t^-b\bar{c}$. But
the contribution of the FC loop-level coupling $tcg$ is embedded
and we can hardly obtain the information of the $tcg$ coupling
from such process.

The decay width and the decay branching ratios of the charged
top-pions have been studied in the references\cite{He, charged
toppion decay}, and the dominant decay modes of $\Pi_t^-$ are
$b\overline{t}$ and $b\overline{c}$. The signal should include two
jets: one is c-jet and another jet includes the particles arising
from $\Pi_t^-$ decaying. The c-jet can be easily identified and
such identification is very important which can help us to confirm
that such production mode is a FC mode and strongly depress the SM
background. To detect $\Pi_t^-$ via the decay mode $b\bar{t}$, the
top quark must be efficiently reconstructed and b-tagging is also
needed. On the other hand, the existence of the FC decay mode
$b\overline{c}$ provides a unique way to detect $\Pi_t^-$. The
decay branching ratio of $b\bar{c}$ is over $10\%$\cite{charged
toppion decay}, so there are enough signals can be produced via
the decay mode $b\overline{c}$ with the yearly luminosity $100
fb^{-1}$ at the LHC. Furthermore, the decay mode $b\overline{c}$
involves the FC coupling $\Pi_t^-b\overline{c}$ which is an
important feature of the TC2 model. So the signal of
$b\overline{c}$ is typical and the SM background is very clean.
Therefore the discovery of the charged top-pions at the LHC would
be possible for a wide range of the parameter space. But the more
detailed study of the background is warranted, in order to
establish the experimental sensitively to the FC coupling
$\Pi_t^-b\overline{c}$. It should be noted that,
 in contrast to the MSSM(Minimal Supersymmetric SM),
 the general 2HDM(Two-higgs Doublet Model) also has potentially the same tree-level FC
couplings for the charged higgs bosons, thus the similar FC
production mode $cH^-$ should exist and the $H^-$ can also decay
to $\bar{t}b$ and $b\bar{c}$. But the difference between the
charged top-pions and charged higgs bosons is that charged higgs
has the extra decay modes, $H^-\rightarrow \tau\nu, \bar{c}s$, and
such difference can help us to distinguish the charged top-pions
from these charged higgs bosons.

\subsection{The $c \Pi_{t}^{0}(h_{t}^{0})$ productions at the hadron colliders}

Besides the charged top-pions, there exist also the neutral CP odd top-pion and CP even top-Higgs
in the TC2 model. The loop-level FC coupling $tcg$ can also induce the FC neutral top-pion and
top-Higgs productions $c\Pi_{t}^{0}(h_t^0)$ at the hadron
colliders. The relevant Feynman diagrams are shown in Fig.2(D-E).

We first study the $c \Pi_{t}^{0}$ production. The amplitudes of
this production mode are expressed as follow
\begin{eqnarray}
M^{c\Pi_t^0}_{D}&=&\frac{-m_{t}^{4}tan^{3}\beta}{16\pi^{2}\upsilon_{w}^{3}}
(1-\varepsilon)(2\varepsilon-\varepsilon^{2})g_{s}T^{a}_{ij}G(p_{1}+p_{2},m_{t})\nonumber\\
&&\{[2B_{1}(-p_{1}-p_{2},m_{b},M_{\Pi_{t}})+B_{1}(-p_{1}-p_{2},m_{t},M_{\Pi_{t}})
+B_{1}(-p_{1}-p_{2},m_{t},M_{h_{t}})\nonumber\\
&&+B_{0}(-p_{1}-p_{2},m_{t},M_{\Pi_{t}})-B_{0}(-p_{1}-p_{2},m_{t},M_{h_{t}})\nonumber\\
&&-B_{0}(-p_{1},m_{t},M_{\Pi_{t}})+B_{0}(-p_{1},m_{t},M_{h_{t}})\nonumber\\
&&-m_{t}^{2}(C'_{0}+C^*_{0})-2m_{b}^{2}C_{0}+4C_{24}+2C'_{24}+2C_{24}^{*}]
\cdot\overline{u}_{c}^{i}(p_{3})\eslash^{a}(p_{2})Ru_{c}^{j}(p_{1})\nonumber\\
&&-[2(C_{23}+C_{12})+C'_{23}+C'_{12}+C^{*}_{23}+C^{*}_{12}]
\cdot\overline{u}_{c}^{i}(p_{3})\pslash_{1}\eslash^{a}(p_{2})\pslash_{2}Ru_{c}^{j}(p_{1})\nonumber\\
&&-(C'_{0}-C^*_{0})
\cdot\overline{u}_{c}^{i}(p_{3})(\pslash_{1}+\pslash_{2})\eslash^{a}(p_{2})\pslash_{2}Ru_{c}^{j}(p_{1})\nonumber\\
&&-(C'_{12}-C^*_{12})
\cdot\overline{u}_{c}^{i}(p_{3})(\pslash_{1}+\pslash_{2})\pslash_{1}\eslash^{a}(p_{2})Ru_{c}^{j}(p_{1})\},
\end{eqnarray}
with the three-point standard functions defined as
$$C_{ij}=C_{ij}(-p_{2},-p_{1},m_{b},m_{b},M_{\Pi_{t}}),$$
$$C'_{ij}=C_{ij}(-p_{2},-p_{1},m_{t},m_{t},M_{\Pi_{t}}),$$
$$C^*_{ij}=C_{ij}(-p_{2},-p_{1},m_{t},m_{t},M_{h_{t}}),$$
and
\begin{eqnarray}
M^{c\Pi_t^0}_{E}&=&\frac{m_{t}^{4}tan\beta^{3}}{16\pi^{2}\upsilon_{w}^{3}}
(1-\varepsilon)(2\varepsilon-\varepsilon^{2})g_{s}T^{a}_{ij}G(p_{3}-p_{2},m_{t})\nonumber\\
&&\{[2B_{1}(p_{2}-p_{3},m_{b},M_{\Pi_{t}})+B_{1}(p_{2}-p_{3},m_{t},M_{\Pi_{t}})
+B_{1}(p_{2}-p_{3},m_{t},M_{h_{t}})\nonumber\\
&&+B_{0}(p_{2}-p_{3},m_{t},M_{\Pi_{t}})-B_{0}(p_{2}-p_{3},m_{t},M_{h_{t}})\nonumber\\
&&-B_{0}(-p_{3},m_{t},M_{\Pi_{t}})+B_{0}(-p_{3},m_{t},M_{h_{t}})\nonumber\\
&&-m_{t}^{2}(C'_{0}+C^*_{0})-2m_{b}^{2}C_{0}+4C_{24}+2C'_{24}+2C_{24}^{*}]
\cdot\overline{u}_{c}^{i}(p_{3})\eslash^{a}(p_{2})Ru_{c}^{j}(p_{1})\nonumber\\
&&+[2(C_{23}+C_{12})+C'_{23}+C'_{12}+C^{*}_{23}+C^{*}_{12}]
\cdot\overline{u}_{c}^{i}(p_{3})\pslash_{2}\eslash^{a}(p_{2})\pslash_{3}Ru_{c}^{j}(p_{1})\nonumber\\
&&+(C'_{0}-C^*_{0})
\cdot\overline{u}_{c}^{i}(p_{3})\pslash_{2}\eslash^{a}(p_{2})(\pslash_{3}-\pslash_{2})Ru_{c}^{j}(p_{1})\nonumber\\
&&+(-C'_{12}+C^*_{12})
\cdot\overline{u}_{c}^{i}(p_{3})\eslash^{a}(p_{2})\pslash_{3}(\pslash_{3}-\pslash_{2})Ru_{c}^{j}(p_{1})\},
\end{eqnarray}
with the three-point standard functions defined as
$$C_{ij}=C_{ij}(p_{2},-p_{3},m_{b},m_{b},M_{\Pi_{t}}),$$
$$C'_{ij}=C_{ij}(p_{2},-p_{3},m_{t},m_{t},M_{\Pi_{t}}),$$
$$C^*_{ij}=C_{ij}(p_{2},-p_{3},m_{t},m_{t},M_{h_{t}}).$$

 To obtain the
numerical results of the $c\Pi_t^0$ production, we choose the same
parameters values as those in the $c\Pi_{t}^{-}$ production. The
cross section of the $c\Pi_{t}^{0}$ production at the Tevatron and
LHC is shown in Fig.5 and Fig.6, respectively.
\begin{figure}[thb]
 \centerline{\mbox{\epsfxsize=8cm\epsffile{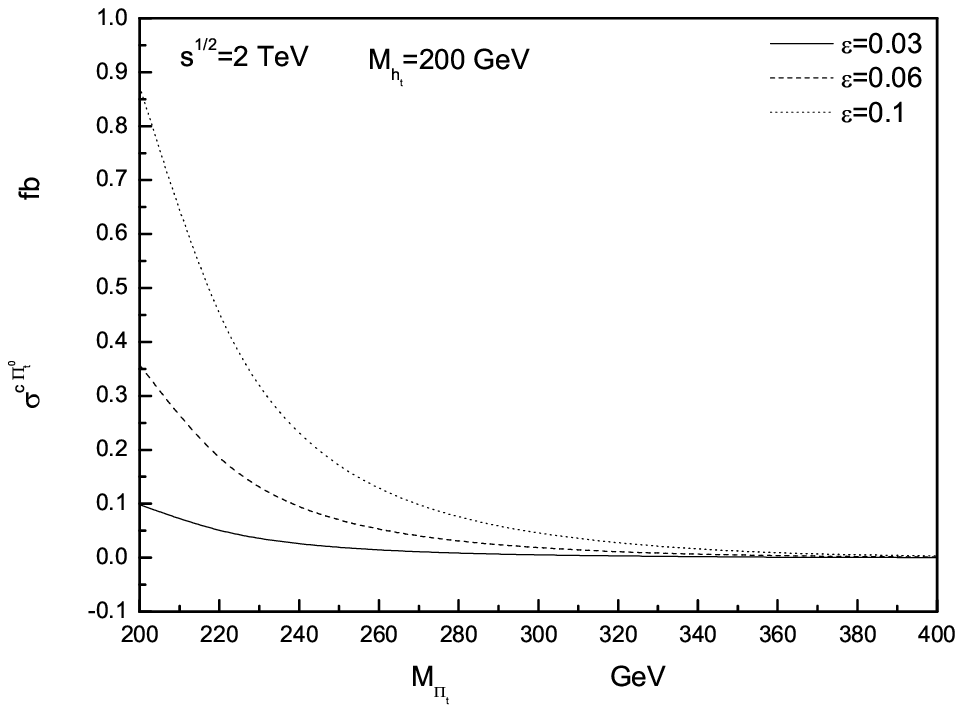}
 \epsfxsize=8cm\epsffile{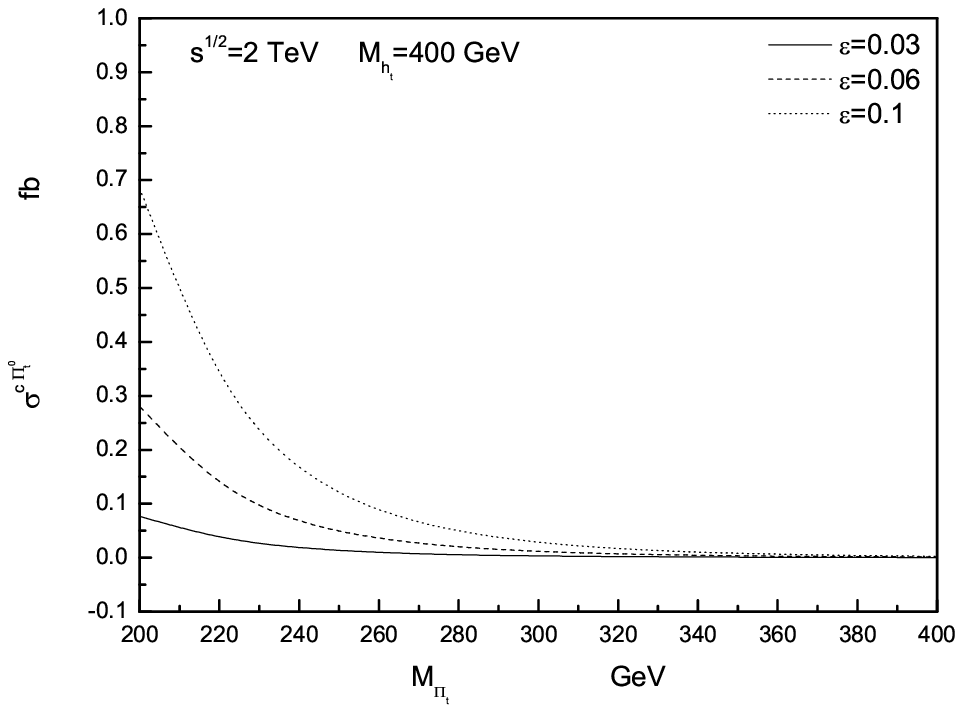}}}
\caption{The hadronic cross section of
$c\Pi_{t}^{0}$ production as a function of $M_{\Pi_t}$ at the
Tevatron, with $M_{h_t}=200, 400$ GeV, respectively.}
\label{fig:fig5}
\end{figure}

\begin{figure}[thb]
 \centerline{\mbox{\epsfxsize=8cm\epsffile{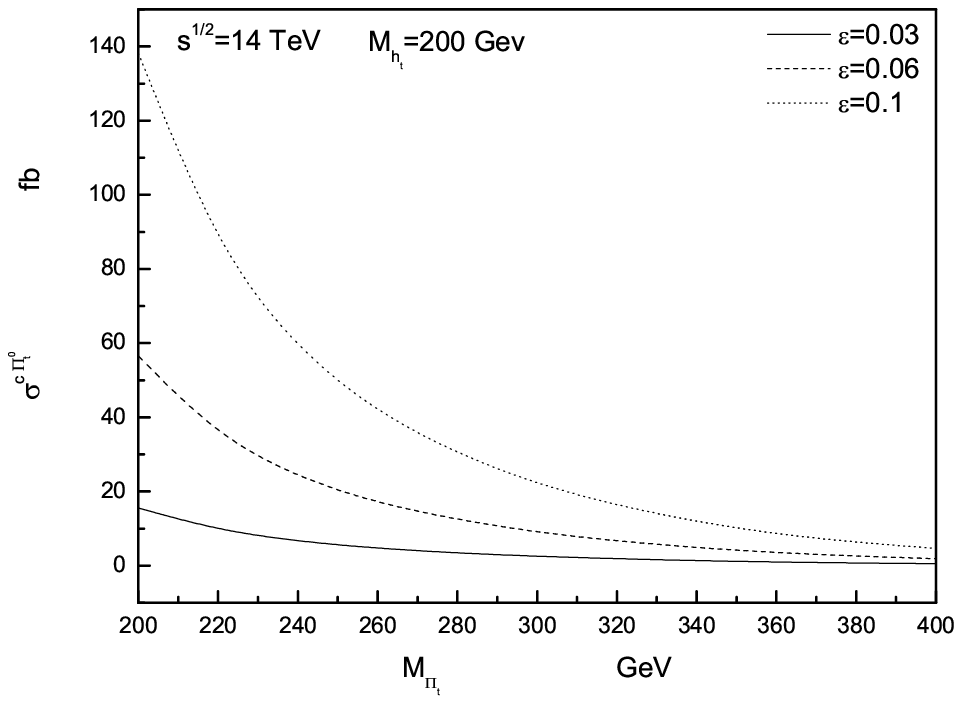}
 \epsfxsize=8cm\epsffile{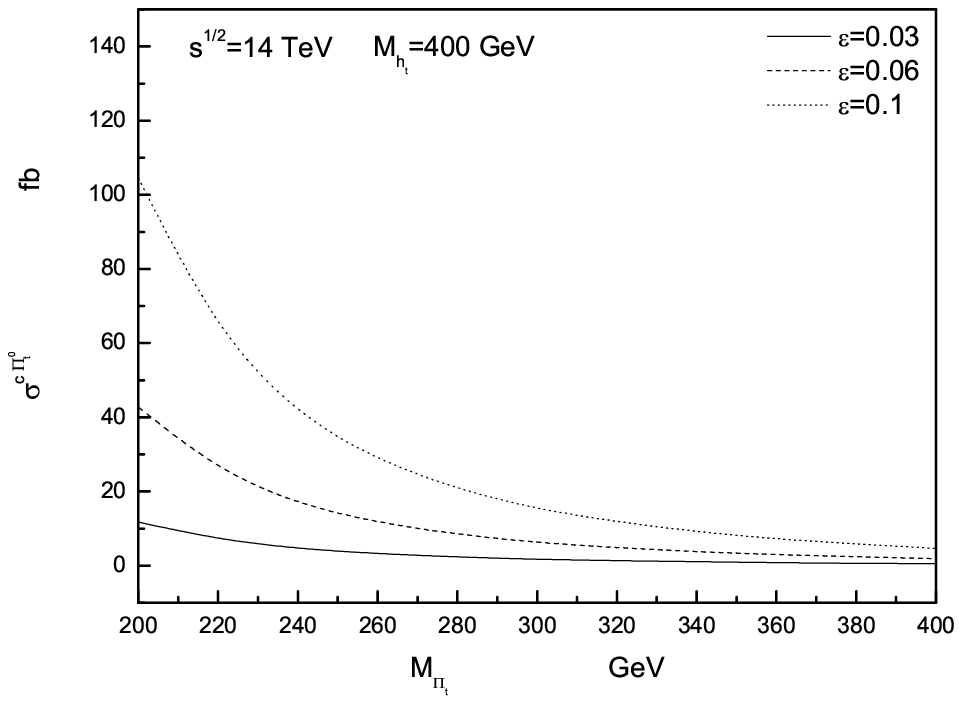}}}
\caption{The hadronic cross section of
$c\Pi_{t}^{0}$ production as a function of $M_{\Pi_t}$ at the LHC,
with $M_{h_t}=200, 400$ GeV, respectively.} \label{fig:fig6}
\end{figure}

As it is shown in Fig.2(D-E), the $c\Pi_t^0$ production only
involves the loop-level FC coupling $tcg$. So the cross section of
the $c\Pi_{t}^{0}$ production is much smaller than that of the
$c\Pi_{t}^{-}$ production. We can see from Fig.5 and Fig.6, the
cross section at the Tevatron is less than one fb which is too
small to detect the neutral top-pion. The cross section at the LHC
can be greatly increased. For the light $\Pi_t^0$, the cross
section is over one hundred fb. There are enough $c\Pi_t^0$
signals would be produced at the LHC in a wide range of parameter
spaces. Therefore, the $c\Pi_t^0$ production at the LHC will open
a good window to search for the neutral top-pion. Furthermore, it
should be noted that the $c\Pi_t^0$ production only involves the
loop-level FC coupling $tcg$. So the $c\Pi_t^0$ production might
also provide a chance to obtain the information of the FC coupling
$tcg$.

The decay branching ratios of the neutral top-pion have been
calculated in the reference\cite{pro-LC2}. The main decay modes of
$\Pi_{t}^{0}$ are $t\overline{t}$, $t\overline{c}$. For heavy
$\Pi_{t}^{0}$($M_{\Pi_{t}}> 2m_{t}$), the channel
$\Pi_{t}^0\rightarrow t\bar{t}$ is open. In this case,
$\Pi_{t}^{0}$ almost decays to $t\bar{t}$ and a large number of
$t\bar{t}$ can the produced. In order to detect $\Pi_t^0$ via
$t\bar{t}$, one should reconstruct the top pair from the final
states and measure the invariant mass distribution of $t\bar{t}$
which make the probe for the $\Pi_{t}^{0}$ via $t\bar{t}$ become
more difficult. So it is a hard work to detect the heavy neutral
top-pion. But the fact that a large number of $t\bar{t}$ events
 associated with a c-jet are produced might provide the clue of the
TC2 model. Below the $t\bar{t}$ threshold, the FC decay channel
$t\bar{c}$ is dominant. Such decay mode involves the typical
feature of the TC2 mode and the peak of the invariant mass
distribution of $t\bar{c}$ is narrow. To identify $t\bar{c}$, one
needs reconstruct top quark from its decay mode $W^+b$.
Furthermore, the b-tagging and c-tagging are also needed. The
experiments can take b-tagging and c-tagging with high
efficiency\cite{efficiency}. So there should be enough clean
$t\bar{c}$ signals for the discovery of $\Pi_t^0$, and the FC
decay mode $t\bar{c}$ is the most ideal one to detect $\Pi_t^0$.
On the other hand, it is also necessary to tag another c-jet
associated with $\Pi_t^0$ production. Such c-tagging can confirm
that the process is a FC process and make the SM background become
very clean.

Now we turn to study the $h_{t}^{0}$ production mode $ch_{t}^{0}$.
For the production amplitudes, the only difference between the
$c\Pi_t^0$ and $ch_t^0$ productions is that there exists a extra
factor $i$ in the $c\Pi_t^0$ production amplitudes, i.e.,
$M^{c\Pi_t^0}_{D}=iM^{ch_t^0}_{D},~
M^{c\Pi_t^0}_{E}=iM^{ch_t^0}_{E}$. So almost identical conclusions
can apply to the $ch_t^0$ production mode, where the only
difference is that there exist the extra tree-level gauge boson
decay modes: $W^+W^-,ZZ$ for $h_t^0$. The decay rates of
$W^+W^-,ZZ$ are suppressed by $r^2(r=m_t/\upsilon_{t})$, but the
branching ratio of $W^+W^-+ZZ$¡¡   can still above $10\%$ if the
decay mode $t\bar{t}$ is forbidden\cite{Burdman}. These gauge
boson decay modes might provide a way to distinguish $h_{t}^{0}$
from $\Pi_t^0$.

\section{Conclusions }
\hspace{1mm}

In this paper, we study the FC production processes of top-pions
and top-Higgs associated with a charm quark. Our study shows that
the cross section of $c\Pi_t^-$ production is much larger than
those of $c\Pi_t^0(h_t^0)$ productions due to the tree-level
contribution of the $b\bar{c} \Pi_t^-$ coupling to the $c\Pi_t^-$
production. At the Tevatron, the cross section is at the order of
tens fb for $c \Pi_t^-$ production and below one fb for $c \Pi_t^0
(h_t^0)$ production in most case. The light charged top-pions are
not favorable by the Tevatron experiments and the Tevatron has a
little capability to probe the neutral top-pion and top-Higgs via
these FC production processes. The cross sections of $c \Pi_t^-$
and $c \Pi_t^0(h_t^0)$ productions can be largely enhanced at the
LHC. The cross sections at the LHC can reach the level $10\sim
100$ pb for $c \Pi_t^-$ production and $10\sim 100$ fb for $c
\Pi_t^0 (h_t^0)$ productions. With the yearly luminosity $100
fb^{-1}$, enough signals could be produced at the LHC. The SM
backgrounds, fortunately, should be clean due to the FC feature of
these processes. Furthermore, the FC decay modes
$\Pi_t^-\rightarrow b\bar{c}$ and $\Pi_t^0(h_t^0)\rightarrow
t\bar{c}$ can provide us the typical signal to detect the
top-pions and top-Higgs. Therefore, it is hopeful to find the
signal of top-pions and top-Higgs with the running of the LHC if
they are there indeed.

\section{Acknowledgments}
\hspace{1mm}

This work is supported  by the National Natural Science Foundation
of China under Grant No.10375017, 10575029 and 10575052.

\end{document}